\documentclass[a4paper,fleqn,usenatbib]{mnras}

\usepackage[T1]{fontenc}
\usepackage{ae,aecompl}
\usepackage{xcolor}

\usepackage{graphicx}	
\usepackage{amssymb}	
\usepackage{pdflscape}	
\usepackage[english]{babel}

\title[The velocity dispersion in molecular clouds]{Clumpy shocks as the driver of velocity dispersion in molecular clouds: the effects of self-gravity and magnetic fields}

\author[D. H. Forgan and I.A. Bonnell]
{D.~H.~Forgan$^1$\thanks{Contact e-mail: \href{mailto:dhf3@st-andrews.ac.uk}{dhf3@st-andrews.ac.uk}},
I.~A.~Bonnell$^1$
\vspace{0.2cm} \\
$^{1}$SUPA, School of Physics \& Astronomy, University of St Andrews, North Haugh, St Andrews, Scotland, KY16 9SS, UK}

\date{Accepted XXX. Received XXX; in original form XXX}

\pubyear{2016}

\begin{document}
\label{firstpage}
\pagerange{\pageref{firstpage}--\pageref{lastpage}}
\maketitle

\begin{abstract}

\noindent We revisit an alternate explanation for the turbulent nature of molecular clouds - namely, that velocity dispersions matching classical predictions of driven turbulence can be generated by the passage of clumpy material through a shock.  While previous work suggested this mechanism can reproduce the observed Larson relation between velocity dispersion and size scale ($\sigma \propto L^{\Gamma}$ with $\Gamma \approx 0.5$), the effects of self-gravity and magnetic fields were not considered.  We run a series of smoothed particle magnetohydrodynamics experiments, passing clumpy gas through a shock in the presence of a combination of self-gravity and magnetic fields.  We find powerlaw relations between $\sigma$ and $L$ throughout, with indices ranging from $\Gamma=0.3-1.2$.  These results are relatively insensitive to the strength and geometry of magnetic fields, provided that the shock is relatively strong.  $\Gamma$ is strongly sensitive to the angle between the gas' bulk velocity and the shock front, and the shock strength (compared to the gravitational boundness of the pre-shock gas).  If the origin of the $\sigma-L$ relation is in clumpy shocks, deviations from the standard Larson relation constrain the strength and behaviour of shocks in spiral galaxies.

\end{abstract}




\section{Introduction}
\label{sec:introduction}

It is well known that giant molecular clouds (GMCs) exhibit highly disordered, chaotic supersonic motions that govern the process of star formation \citep[see e.g.][ for a review]{McKee2007}.  Generally speaking, these motions show evidence of energy injected into a range of size scales, a necessary characteristic of turbulence \citep{Elmegreen2004}.

The evidence for supersonic turbulence is boosted by the observed power-law relationship between the velocity dispersion in the gas $\sigma$ at a given size scale $L$:
\begin{equation}
\sigma \propto L^{\Gamma},
\end{equation}


\noindent where $\Gamma \approx 0.5$ \citep{Larson1981}.  The universal proportionality of this relationship is supported by a series of observations of CO linewidths \citep{Heyer2004, Rice2016,Sun2017}, and interpreted observations of dust polarisation \citep[e.g.][]{Poidevin2013}.  While the proportionality appears to be universal, the magnitude of the velocity dispersion does increase with increasing surface density \citep{Heyer2009}.    This behaviour is demonstrated at scales ranging from 0.01 pc to tens of parsecs, with obvious exceptions being regions where gravitational collapse dominates the evolution, such as collapsing prestellar cores (see also \citealt{Leroy2016}, who show that higher surface density regions tend to exhibit higher linewidths for a given $L$).

This relatively universal $\sigma-L$ relation suggests that the ISM is indeed turbulent, with a large-scale driving source in operation \citep{Kritsuk2013}. Indeed, most forms of classical turbulence predict a $\sigma-L$ relation of some form, be it Kolmogorov turbulence \citep[$\Gamma=0.33$,][]{Passot1988}, Burger's shock-driven turbulence \citep[$\Gamma=0.5$,][]{Scalo1998}, or She-Leveque turbulence \citep[$\Gamma=0.42$,][]{She1994, Boldyrev2002}.  For any of these relations to hold, energy injection must occur on scales upwards of several hundred parsecs. This is in accord with synthetic observations of simulated molecular clouds, which are only consistent with ``real'' observations when the turbulence is being driven at large scales \citep{Brunt2009}.

What is the large-scale driving source? There is a large list of possible agents.  Radiative and hydrodynamic feedback from star formation is one possibility.  Winds, ionisation fronts and particularly supernovae can inject significant quantities of kinetic energy into the local environment \citep{Gressel2008}.  Interaction with the magnetic field can then generate magnetohydrodynamic (MHD) turbulence, which simulations indicate can successfully reproduce the statistical behaviour of ISM gas \citep{Kritsuk2007,Federrath2010,Padoan2011}.

It is more challenging to argue that this feedback can act universally, as it is typically only active where star formation itself is active.  It is also the case that the efficacy of supernova feedback becomes sensitive to the age (i.e. density) of the molecular cloud, as well as the distance of the supernova to the cloud \citep{Ibanez-Mejia2017,Seifried2018}.

It is true that feedback can trigger star formation \citep[e.g.][]{Bisbas2011}, but there are also examples of where feedback inhibits star formation \citep{Lucas2017}.   Typically, feedback provides a mix of triggering and inhibition depending on the cloud's density structure and the distribution of feedback agents \citep{Dale2007a,Dale2014}.  Also, radiative feedback generally acts asymmetrically, with most energy injection occurring perpendicular to the galactic disc, where optical depths at long range are typically lowest \citep{Henley2010}.

Turbulence can be produced directly by instabilities in the disc gas.  The galactic disc will become gravitationally unstable if the Toomre parameter 

\begin{equation}
Q = \frac{c_s \kappa }{\pi G \Sigma} \approx 1.5-1.7
\end{equation}

\noindent where $c_s$ is the sound speed of the gas, $\kappa$ is the epicyclic frequency and $\Sigma$ is the gas surface density. This can drive internal motions on scales of order tens to hundreds of parsecs \citep[see e.g.][]{Goldbaum2015}.  While a promising agent for a global driver of turbulence, gravitational instability tends to act more vigorously in the regions where the gas has been able to cool efficiently, i.e. once it is already neutral and molecular.  The gas's low temperature tends to result in low velocity turbulence, and the motions of the gas are typically quite coherent as opposed to chaotic (although this can vary depending on the geometry of gas inflow/accretion).  Other instabilities, such as the magneto-rotational instability, have been proposed to produce velocity dispersion, again with a strong dependence on accretion \citep{Klessen2010}.

Converging/colliding flows can also drive strong internal motions.  The shocks produced by these collisions can result in efficient cooling of the post-shock gas, giving rise to dense, cool layers which can result in fragmentation and complex velocity fields \citep{BallesterosParedes1999,Heitsch2006,Wu2018}.  

Spiral shocks efficiently produce converging flows, as matter flows into the spiral arm potential.  Warm, low density gas flowing into the arm region is compressed and shocked, dissipating kinetic energy and permitting the formation of high density regions that can then collapse under gravity, even if the resulting molecular clouds are themselves gravitationally unbound \citep{Bonnell2006}, with star formation rates following the Schmidt-Kennicutt relation ($\Sigma_{\rm SFR} \propto \Sigma_{\rm gas}^{1.4}$), and a $\sigma-L$ relation with $\Gamma = 0.5$ \citep{Bonnell2013} .   

\citet{Dobbs2007} conducted a series of numerical experiments to demonstrate that a $\sigma-L$ relation can be generated in the passing of clumpy material through a shock (generated by a spiral arm for example).  This is despite the material initially possessing an entirely uniform bulk (supersonic) velocity.  They interpret this as a consequence of the ``mass-loading'' experienced by the gas  while passing through the shock.  The mass loading is a function of size scale - small size scales typically encounter regions of similar column density during the shock, and hence the gas exhibits low velocity dispersion.  As the size scale increases, the region samples a wide range of column densities during the shock, and as a result exhibits a wide range of post-shock velocities, and thus a high velocity dispersion.

This ``clumpy shock'' origin of ISM turbulence can be seen as a combination of both gravitational instability and convergent flows, where GI forms the spiral arm that induces the shock, and the velocity differential in the gas that permits converging flows (\citealt{Bonnell2013}, see also \citealt{Falceta-Goncalves2014}).  In their work \citet{Dobbs2007} simulated the flow using smoothed particle hydrodynamics (SPH) without the effects of self-gravity.   They also omit magnetic fields from their analysis. 

Clearly these forces will shape the resultant velocity field of molecular clouds, and may significantly alter the value of $\Gamma$ produced by clumpy shocked material.  We revisit these calculations using smoothed particle magnetohydrodynamics (SPMHD), and explore the effects of both self-gravity and magnetic field on the resulting $\sigma-L$ relation.


\section{Method}
\label{sec:method}

\subsection{Phantom}

\noindent Smoothed Particle Hydrodynamics (SPH) is a Lagrangian method for solving the equations of fluid dynamics.  The fluid is decomposed into a collection of particles, each possessing a mass $m_i$, position $\mathbf{r}_i$, velocity $\mathbf{v}_i$, internal energy $u_i$ and smoothing length $h_i$ (where $i$ is the particle label). The density of the fluid at any position is reconstructed using the kernel weighted estimator

\begin{equation}
\rho (\mathbf{r}) = \sum^{N}_{i=1} m_i W(\left|\mathbf{r} - \mathbf{r}_i\right|, h),
\end{equation}
where $W$ is the smoothing kernel.  The kernel function is selected to have compact support within the range  $\left|\mathbf{r} - \mathbf{r}_i\right| = [0,2h]$, so that $N$ represents the number of neighbouring particles within a distance $2h$ of $\mathbf{r}$.

The equations of motion for the fluid proceed entirely from this density estimator (with an appropriate Lagrangian and variational principle), yielding a consistent framework for solving the (magneto)hydrodynamic equations (see \citealt{Price2012} for a review).  We use the SPMHD code \textsc{Phantom} \citep{Price2017}. The implementation of MHD in \textsc{Phantom} follows the basic scheme described in \citet{Price2004,Price2004a,Price2005a} (see review by \citealt{Price2012}) with the divergence constraint on the magnetic field enforced using the constrained hyperbolic divergence cleaning algorithm described by \citet{Tricco2012} and \citet{Tricco2016}. We assume ideal MHD for this work.

We employ artificial viscosity, conductivity and resistivity to resolve shocks and prevent unphysical particle interpenetration, for viscosity adopting the time-dependent viscosity of \citet{Morris1997}, where the $\Gamma_{SPH}$ can vary between 0.1 and 1, and the corresponding non-linear viscosity term is fixed at $\beta_{SPH}=2$. The particles evolve on individual timesteps, and the gravity forces are computed using a binary tree similar to that described in \citet{Gafton2011}.

The gas equation of state is adiabatic, with ratio of specific heats $\gamma=5/3$.  We do not impose any radiative cooling prescription on the gas.  Sink creation is prohibited.  

\subsection{Initial Conditions}

\noindent Unlike \citet{Dobbs2007}, we are most interested in the effect of varying the physical forces in play, rather than varying the clumpiness or other properties of the gas.  Therefore, all runs have the same initial conditions.  We initialise a clumpy box of gas using 400000 particles, each with particle mass $3.29\times10^{-5}$.  The box has dimensions of $-1.5 < x < 1.5$, $-1.5 < y < 1.5$, and $-1 < z < 1$.  Half of the box by mass is composed of uniform density gas, with the other half composed of spherical clumps, distributed randomly through the box.  Each sphere has a fixed radius of $r_{\rm cl} = 0.1$, and a density 30 times the mean box density.  The clumps are initially Jeans stable, with a free fall time around ten times the shock crossing time across the clump.  This ensures that generation of velocity dispersion will not proceed due to the self-gravitating collapse of the clumps, but how the clumpy gas enters and leaves the shock (see section 3.3 of \citealt{Dobbs2007}).

The gas has initially zero motion in the $y$ and $z$ direction.  The x-velocity is uniform across the gas, with $v_x=50 c_s$, where the sound speed of the gas is $c_s=0.19$ code units.

The box is initially centred on the origin of the co-ordinate system.  As it moves in the positive $x$-direction, it encounters a sinusoidal external potential, designed to mimic the passage of the gas through a spiral shock.  The external potential is given by 

\begin{equation}
\phi = A \cos k \left(x + B\right)
\end{equation}

\noindent We fix $k=\pi/4$, and $B=2.0$, to ensure a potential minimum at $x=2$.  In the majority of cases, we fix $A=100$, although we do try some runs with a weaker shock ($A=10$).

Table \ref{tab:purehydro} displays the non-MHD runs carried out in this work.  These are designed to investigate the effects of self-gravity of the gas (SG), and the strength of the shock (CW, SGW).  We also investigate the effect of shock orientation to the bulk flow of the gas (C45).

\begin{table}
\centering
  \caption{The non-MHD simulations carried out \label{tab:purehydro}}
  \begin{tabular}{c | c}
  \hline
  \hline
   Simulation  &  Properties  \\
   \hline
   C & Control - Pure hydro, no self-gravity, $A=100$ \\ 
   C45 & Pure hydro, no self-gravity, $A=100$, 45$^\circ$ shock \\
   CW & Pure hydro, no self-gravity, $A=10$ \\
   SG & Pure hydro, self-gravity, $A=100$ \\
   SGW & Pure hydro, self-gravity, $A=10$ \\
 \hline
  \hline
\end{tabular}
\end{table}

\begin{table}
\centering
  \caption{The ideal MHD simulations carried out in this paper \label{tab:MHD}}
  \begin{tabular}{c | c}
  \hline
  \hline
   Simulation  &  Properties  \\
   \hline
   B100A & $\beta_p=100$, field parallel to the shock \\
   B100P & $\beta_p=100$, field perpendicular to the shock \\
   B100-45 & $\beta_p=100$, field 45$^\circ$ to the shock \\
   B100A-S45 & $\beta_p=100$, field parallel to the y-axis, shock at 45$^\circ$ \\
   B1A & $\beta_p=1$, field parallel to the shock \\
   B1P & $\beta_p=1$, field perp. to the shock \\
   B100ASG & $\beta_p=100$, field parallel to the shock, self-gravity \\
   B100PSG & $\beta_p=100$, field perp. to the shock, self-gravity \\
   B1ASG & $\beta_p=1$, field parallel to the shock, self-gravity \\
   B1PSG & $\beta_p=1$, field perp. to the shock, self-gravity\\
 \hline
  \hline
\end{tabular}
\end{table}

\noindent We initialise MHD simulations with a constant plasma $\beta_p$, i.e. the magnetic pressure at any point is initially a fixed ratio of the local pressure.  There are three factors whose dependence we must now test - the velocity of the gas, the shock, and the magnetic field itself.  Table \ref{tab:MHD} shows the various MHD parameters investigated, which include comparisons with and without self-gravity.  To investigate orientation effects, we conduct a run where the B-field is 45$^\circ$ to the bulk flow and the head-on shock (B10045), and a run where the B-field is parallel to the y-axis (i.e. perpendicular to the flow), and the shock is 45$^\circ$ to the velocity (B100A-S45).

For each MHD run we calculate 

\begin{equation}
\frac{h \nabla .\mathbf{B}}{\left| \mathbf{B} \right|}
\end{equation}

\noindent for all the particles to monitor the SPMHD algorithm's divergence cleaning performance.  We find that for all the simulations run in this paper, the mean value of this quantity is below 0.1.  Isolated particles do occasionally exceed unity but are rare.

\subsection{Calculating velocity dispersion - size scale relations}

\noindent As SPH is a Lagrangian method, and each particle possesses a velocity equal to the fluid velocity at that location we are free to calculate velocity dispersions for collections of fluid elements (i.e. particles).  For a given size scale $L$, we calculate $\sigma$ as follows.

Beginning at the densest particle $j$, we calculate $\sigma_j$ from the velocities of the particles $\{i\}$ contained within the sphere defined by $\left|\mathbf{r}_i - \mathbf{r}_j\right| < L$.  This is then repeated for the next most dense particle, ensuring that particles that have participated in previous sums/calculations are not double-counted.  We do not impose any minimum density limits for this calculation (although see section \ref{sec:density}).

This gives a set of $N_{\sigma_j}$ velocity dispersions $\{\sigma_j\}$ for this $L$.  We then compute the expectation and variance of this set:

\begin{equation}
E(\sigma_j) \equiv \sigma(L) = \frac{1}{N_{\sigma_j}} \sum_j \sigma_j.
\end{equation} 

The variance then gives us an error estimate for $\sigma(L)$:

\begin{equation}
V(\sigma_j) = \frac{1}{N_{\sigma_j}-1} \sum_j (\sigma_j - E(\sigma_j))^2.
\end{equation}

\noindent In the plots of $\sigma(L)$ that follow, we plot $\sqrt{V(\sigma_j)}$ as error bars.

\section{Results}
\label{sec:results}

\subsection{Control Run}

\begin{figure*}
\begin{center}
$\begin{array}{cc}
\includegraphics[scale=0.25]{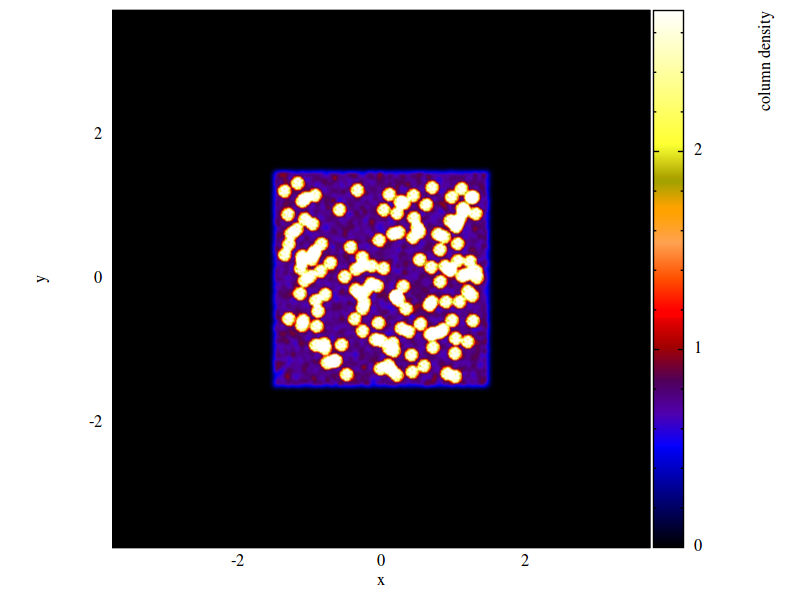} &
\includegraphics[scale=0.25]{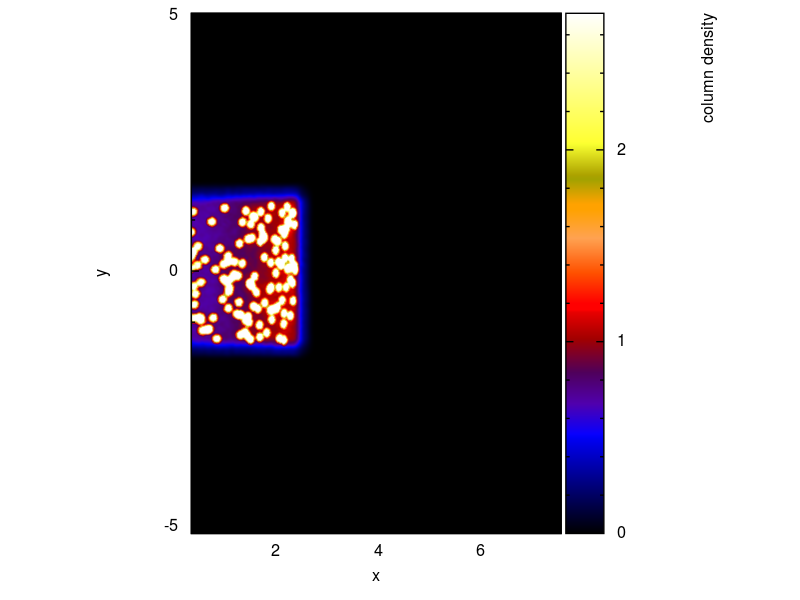} \\
\includegraphics[scale=0.25]{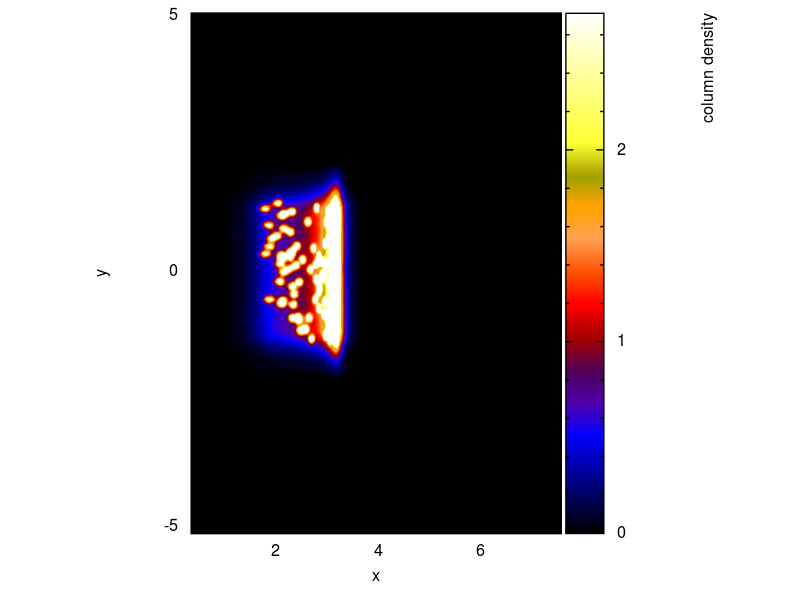} &
\includegraphics[scale=0.25]{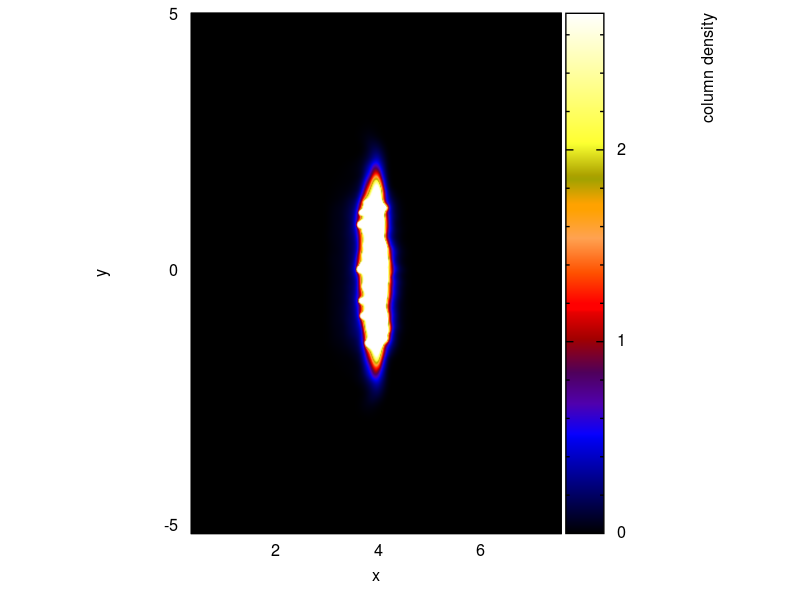} \\
\includegraphics[scale=0.25]{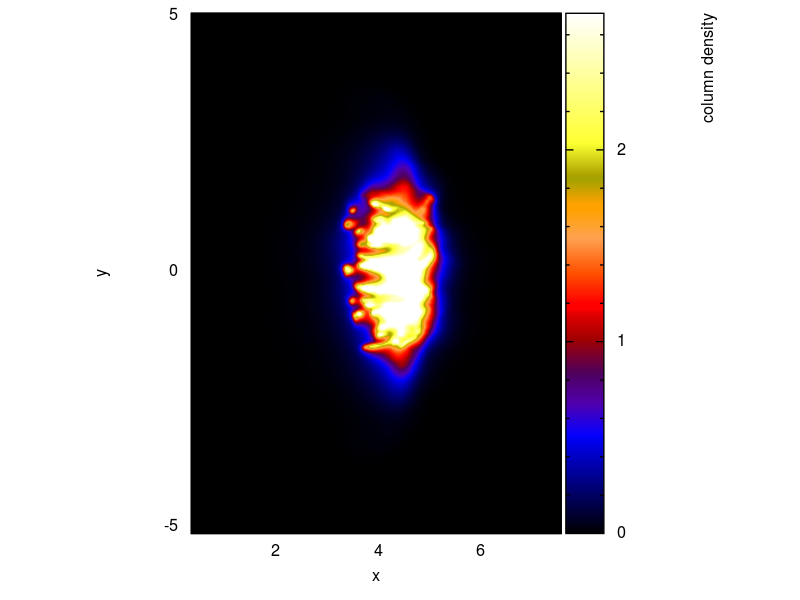} &
\includegraphics[scale=0.25]{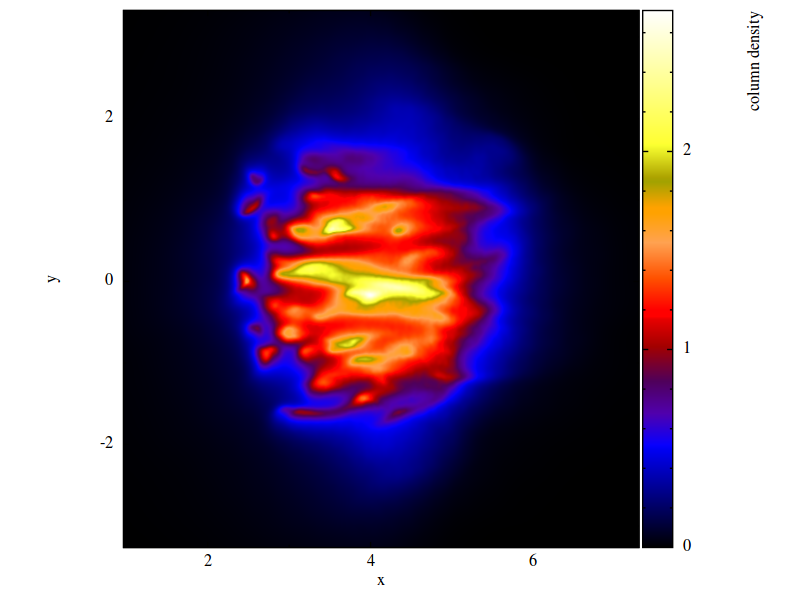} \\
\end{array}$
\end{center}
\caption{The evolution of the clumpy box of gas as it passes through the sinusoidal shock in the control simulation (C).  The initial conditions are shown in the top left panel.  The shock is parallel to the y-axis, located at $x=2$. \label{fig:control_density}}
\end{figure*}

\begin{figure}
\begin{center}
\includegraphics[scale=0.4]{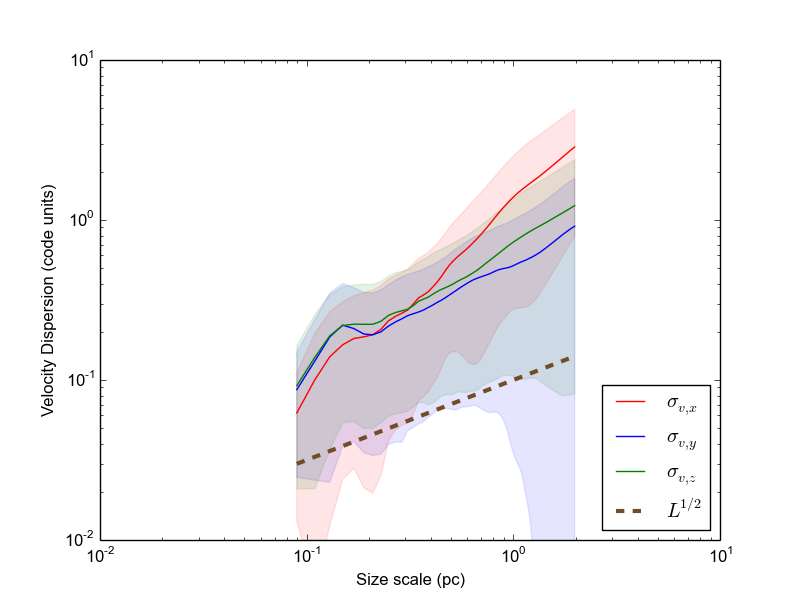}
\end{center}
\caption{The velocity dispersion as a function of size scale for the $x,y,z$ components of the control run, measured at $t=0.5$ units.  In this plot and similar plots throughout this paper, we also add a representative $L^{1/2}$ line for reference. \label{fig:control_3vdisp}}
\end{figure}

\noindent Figure \ref{fig:control_density} shows the evolution of the gas density in the control run (C) from its initial condition (left panel), to after its passage through the sinusoidal shock (right panel).  The initial clumpy structure (left) has a uniform bulk velocity in $x$, and no velocity in $y$ and $z$. This clumpy structure is soon erased by the compressive action of the shock.  The subsequent motions generated result in velocity dispersions along all three co-ordinate axes (Figure \ref{fig:control_3vdisp}).  

As can be seen, the $x$-velocity (red line in Figure \ref{fig:control_3vdisp}) generates the strongest dispersion, especially at large scales (note that the largest initial scale of the box is $4$ pc).  At scales below approximately 0.3 pc, all three velocity components show similar dispersions, and a similar scaling with $L$.  

The scatter in each component is significant (as indicated by the shaded regions around each curve).  We attempt to fit the $\sigma-L$ relation in log-space

\begin{equation}
\log \sigma = \Gamma \log L + C
\end{equation}

\noindent And find that the best-fit $\Gamma=1.2$ for $x$, and $\Gamma \approx 0.7$ for the $y$ and $z$ components.  Of course, the significant scatter indicates that there exists a wide range of values for $\Gamma$ with relatively low $\chi^2$.  

Nonetheless, this highlights a common trend, which was also noted by \citet{Dobbs2007}: if the material arrives at the shock head-on, the velocity dispersion along that axis tends to be boosted compared to the other axes.

\subsubsection{The evolution of the $\sigma-L$ relation}

\begin{figure}
\begin{center}
\includegraphics[scale=0.4]{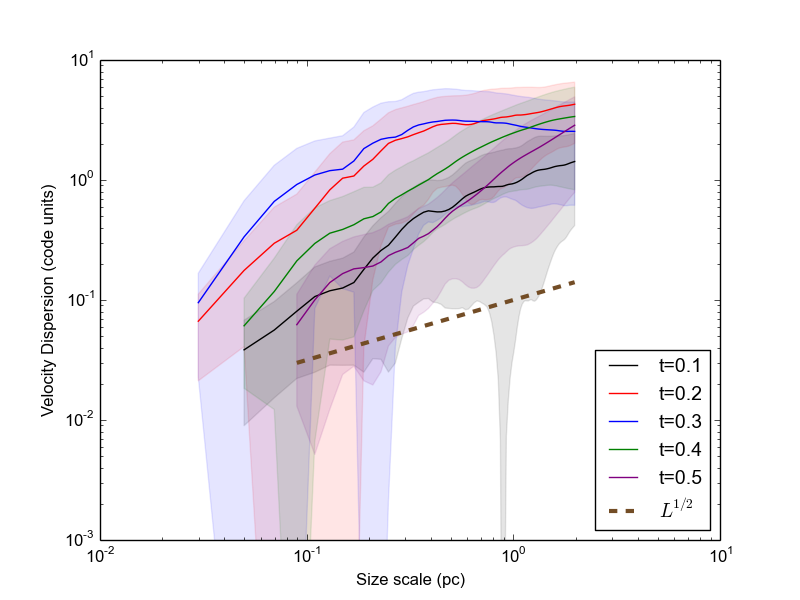}
\end{center}
\caption{The evolution of the $\sigma-L$ relation in the control run over several timesteps.  The curves correspond to the five snapshots plotted in Figure \ref{fig:control_density}. \label{fig:control_multistep}}
\end{figure}

\noindent Figure \ref{fig:control_multistep} shows the evolution of the $\sigma-L$ relation in the control run over five snapshots.  In the first ($t=0.1$ units) the front edge of the box begins to enter the shock, resulting in a compression of the ambient gas (without appreciable effect on the clumps contained in the box).  By $t=0.2$ units, most of the ambient gas is now being compressed, with a few clumps still retaining their structure.  The instant of maximum compression occurs at $t=0.3$ units, with the final two images showing the gas' post-shock behaviour. 

\begin{figure}
\begin{center}
\includegraphics[scale=0.4]{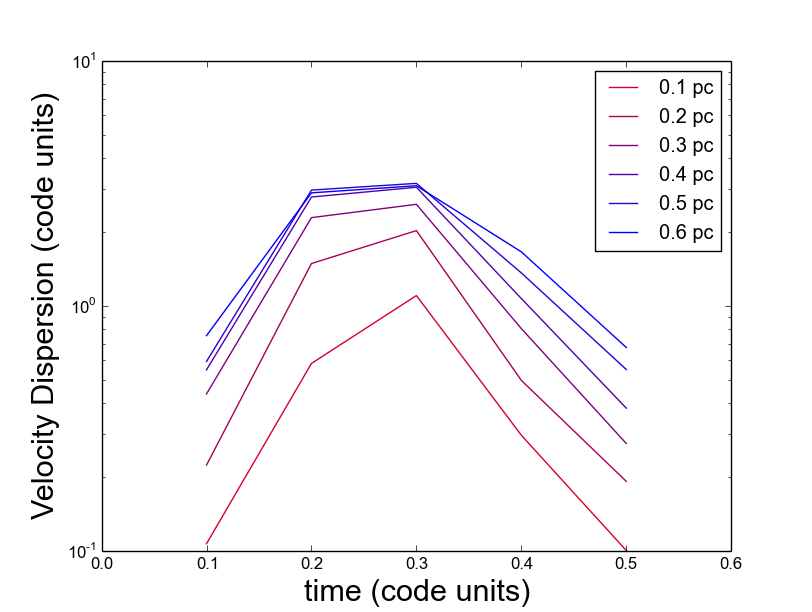}
\end{center}
\caption{The time evolution of $\sigma$ at a series of size scales \label{fig:control_sigmatime}}
\end{figure}

We can see from Figure \ref{fig:control_multistep} that even at a relatively early stage, a powerlaw relation is being set up in the gas that is interacting with the shock.  The velocity dispersion at scales larger than $r_{cl}=0.1$ pc quickly grow, so that the dispersion relation flattens during the instant of maximum compression (see also Figure \ref{fig:control_sigmatime}).  As the gas leaves the shock, this begins to settle back towards its earlier relation.  Over the course of the box's entry and exit from the shock, the value of $\Gamma$ (for $v_x$) varies between 0.5 at early times, reducing to 0.37 during maximum compression, and then increasing back to 1.2 at late times.  

\subsubsection{Density dependence \label{sec:density}}

\noindent When attempting to observe configurations of shocked material such as those seen here, we rely on molecular line tracers, which only emit above a critical density.  In the above plots, we have considered the contribution of all gas to the $\sigma-L$ relation.  How does this relationship change if we only consider the most dense material?

We recalculated the $\sigma-L$ relation for both the C and C45 runs, only counting contributions from the particles above the 30th percentile of density.  There was no appreciable difference to the resulting relations, or the fits to $\Gamma$ in $v_x$, $v_y$ or $v_z$.  Importantly, there was no appreciable change in the scatter derived in this relation.

\subsubsection{Dependence on shock entry angle}

\begin{figure*}
\begin{center}$\begin{array}{cc}
\includegraphics[scale=0.4]{figs/allvdisp_control.png} &
\includegraphics[scale=0.4]{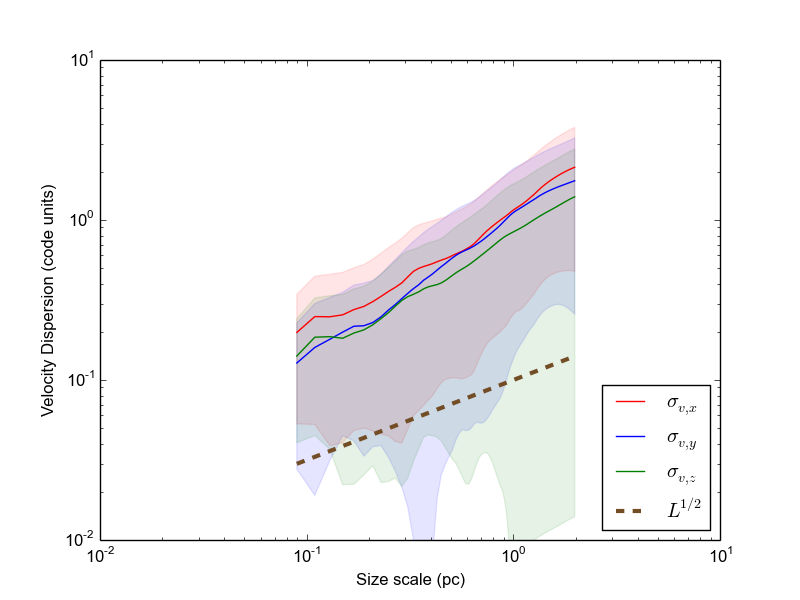} \\
\end{array}$
\end{center}
\caption{The velocity dispersion as a function of size scale for the $x,y,z$ components as a function of shock angle.  Left: the control run (C); right, the control run with shock arriving at a 45$^\circ$ angle (C45).  Both are measured at $t=0.5$ units. \label{fig:control_angle}}
\end{figure*}

\noindent Figure \ref{fig:control_angle} shows the change in behaviour as the shock front is rotated by 45$^\circ$ around the $z$-axis.  Unsurprisingly, the velocity dispersion in the $y$ direction increases.  The velocity dispersion in $z$ remains unchanged.  As a result, the difference in best-fit $\Gamma$ for all co-ordinates is decreased (Table \ref{tab:hydro_fits}). 

\subsection{Pure hydrodynamic runs}

\begin{figure*}
\begin{center}
$\begin{array}{c}
\includegraphics[scale=0.5]{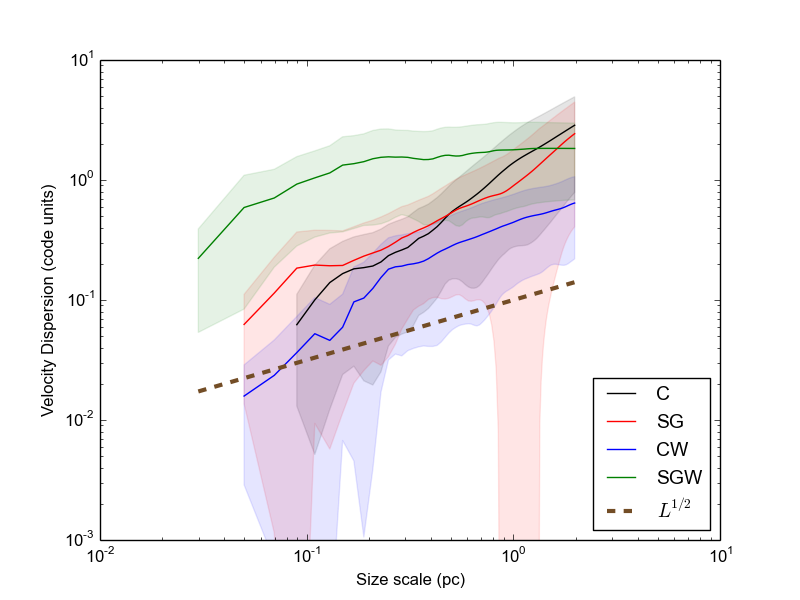} \\
\includegraphics[scale=0.5]{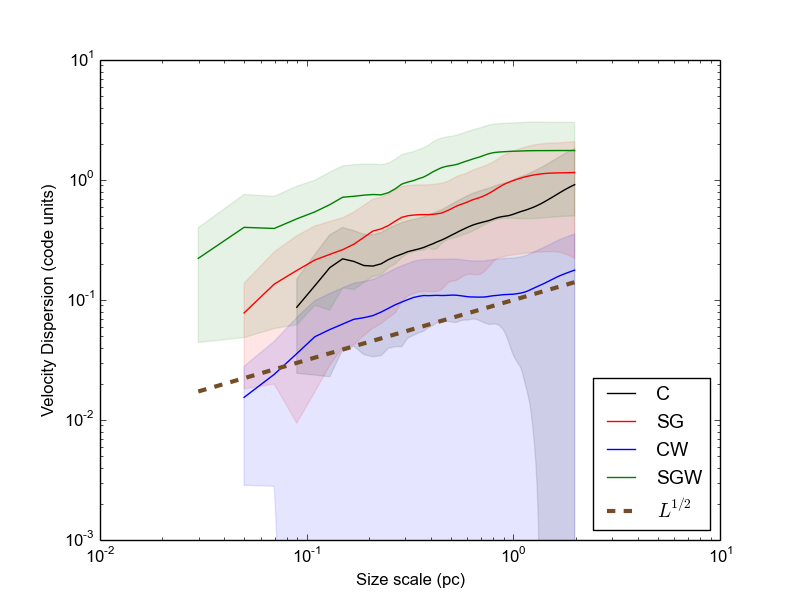} \\
\includegraphics[scale=0.5]{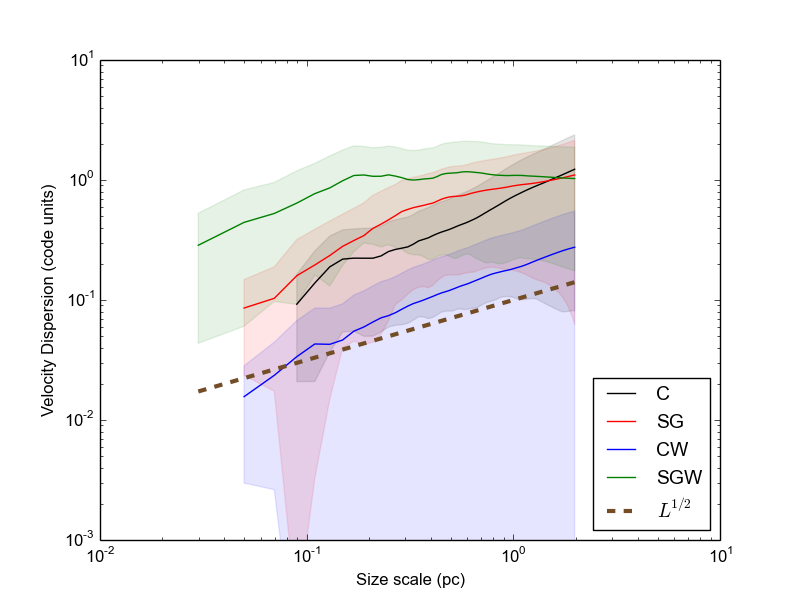} \\
\end{array}$
\end{center}
\caption{The velocity dispersion as a function of size scale in the post-shock gas for the $x$ component (top) $y$ component (middle) and $z$ component (bottom), for the pure hydrodynamic simulations.  All are measured at $t=0.5$ units. }
\label{fig:hydro_runs}
\end{figure*}

\begin{table}
\centering
  \caption{The best-fit $\Gamma$ for the non-MHD simulations \label{tab:hydro_fits}. Recall W=weak shocks, SG=self-gravity.}
  \begin{tabular}{c | ccc}
  \hline
  \hline
   Simulation  &  $\Gamma \,(v_x)$ & $\Gamma \,(v_y)$& $\Gamma \,(v_z)$ \\
   \hline
   C & 1.2 & 0.7 & 0.77 \\
   C45 & 0.952 & 0.840 & 0.788 \\
   CW & 0.69 & 0.37 & 0.64 \\
   SG & 1.11 & 0.335 & 0.35 \\
   SGW & 0.122 & 0.3 & -0.03 \\
 \hline
  \hline
\end{tabular}
\end{table}

\noindent We plot the $x,y$ and $z$ components of the velocity dispersion as a function of size scale for the various pure hydrodynamic runs in Figure \ref{fig:hydro_runs}.  We can see that the addition of self-gravity tends to slightly reduce $\Gamma$ in the direction of travel ($x$), with a more pronounced reduction in the $y$ and $z$ co-ordinates (see also Table \ref{tab:hydro_fits}, which shows our fitted values for each run/component).

As might be expected, reducing the shock strength by a factor of ten also reduces the resulting velocity dispersions generated, as can be seen by comparing the C and CW runs.  The amplitude of $\sigma$ decreases at all scales by around a factor of 2, and the resulting $\Gamma$ is also generally lower.  This becomes quite extreme if self-gravity is activated.  The weakness of the shock delivers correspondingly weak compression of the clumpy gas.  As a result, the post shock gas retains more of its original clumpy structure, and the ensuing velocity dispersion is being generated by the gravitational collapse of this compressed material.

Reducing the shock amplitude $A$ with self-gravity active (SGW) tends to flatten the relation, resulting in very low $\Gamma$ values.  A similar result is achieved if $A$ is fixed at its control value, but the total mass of the shocked material is changed.  This indicates that $\Gamma$ is controlled by the relationship between the boundness of the gas pre-shock and the depth of the spiral potential.  This is consistent with the emerging picture from global simulations of molecular cloud formation in spiral structures \citep[e.g.][]{Jin2017}, where molecular cloud turbulence parameters assume a spectrum of values depending on the details of their shock ingress/egress.

\subsection{MHD runs without self-gravity}

\begin{figure*}
\begin{center}
$\begin{array}{c}
\includegraphics[scale=0.5]{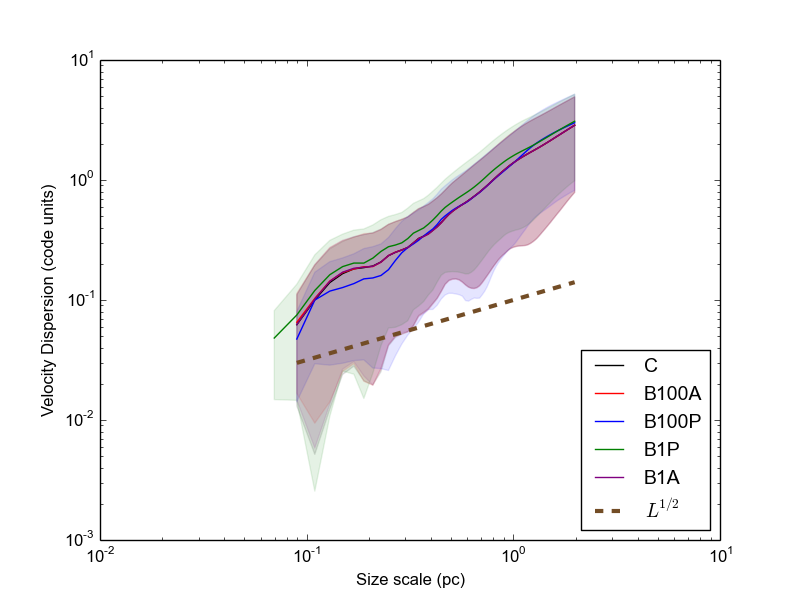} \\
\includegraphics[scale=0.5]{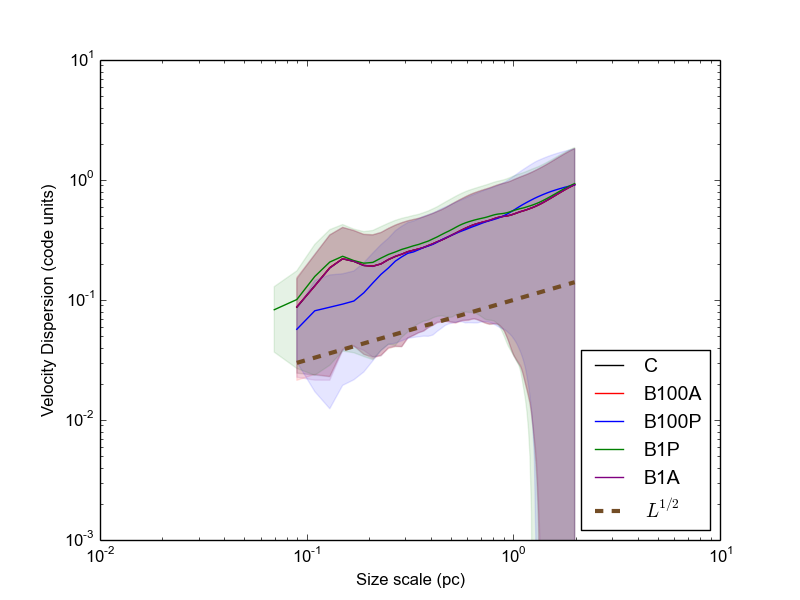} \\
\includegraphics[scale=0.5]{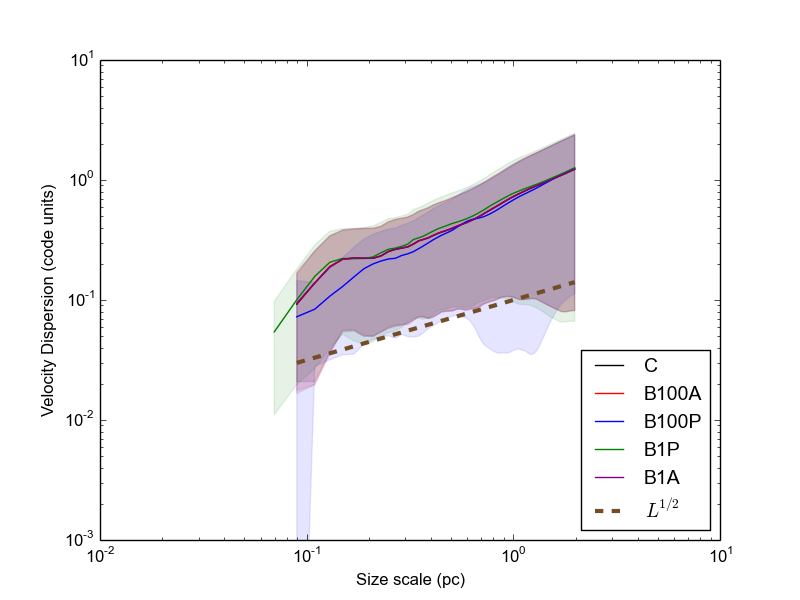} \\
\end{array}$
\end{center}
\caption{The velocity dispersion as a function of size scale for the $x$ component (top) $y$ component (middle) and $z$ component (bottom), for the ideal MHD simulations.  All are measured at $t=0.5$ units.}
\label{fig:MHD_runs}
\end{figure*}

\begin{table}
\centering
  \caption{The best-fit $\Gamma$ for the non-self-gravitating MHD simulations \label{tab:MHD_fits}}
  \begin{tabular}{c | ccc}
  \hline
  \hline
   Simulation  &  $\Gamma \,(v_x)$ & $\Gamma \,(v_y)$& $\Gamma \,(v_z)$ \\
   \hline
   B100A & 1.2 & 0.74 & 0.77 \\
   B100P  & 1.36 & 0.703 & 0.899 \\
   B1A & 1.136 & 0.701 & 0.744 \\
   B1P  & 1.208 & 0.74 & 0.77 \\
 \hline
  \hline
\end{tabular}
\end{table}

\noindent Figure \ref{fig:MHD_runs} shows the velocity dispersion in all three co-ordinates for MHD runs without self-gravity.  Regardless of the magnetic field strength or orientation, we find that the resulting $\sigma-L$ relations are similar to the control run.  This suggests that magnetic pressure is playing a limited role in resisting gas compression, even when the magnetic pressure and thermal pressure are initially equal ($\beta_p=1$).  This is somewhat in contrast with local MHD simulations with externally forced turbulence, where the resulting turbulence and cloud properties are sensitive to the Alfv\'{e}n Mach number \citep{Padoan1999}.  It is worth noting that our simulations are only subject to a single forcing event, and multiple forcing events are required to produce genuine MHD turbulence, so we should take care when comparing our work to the literature in this respect.  That being said, the $\sigma-L$ relation produced in externally forced turbulent simulations appears to be insensitive to the nature of the turbulence, either compressive or solenoidal \citep{Kritsuk2007,Federrath2010}.  This model is to some degree in agreement with this. 

Our results suggest that the depth of the potential is sufficiently large to make ideal MHD effects irrelevant in the generation of a velocity dispersion relation through spiral shocks.  To check this, we repeat the B1A simulation, but reduce the shock strength $A$ from 100 to 10.  We find that the resulting best fit $\Gamma$ profiles are similar to the same simulation without magnetic fields (CW), but that velocity dispersion in $x$ becomes flat beyond $r>0.3$ pc, at a value of around 0.2 code units, as compression in the shock is now being effectively resisted by magnetic pressure, as expected.  This therefore indicates that in this model, MHD effects are only important when the spiral shock becomes relatively weak\footnote{Even if MHD effects are not important in the generation of velocity dispersion, that of course does not rule out their importance in regulating star formation \citep[cf][]{Padoan2011,Federrath2016}}.

\subsection{MHD runs with self-gravity}

\begin{figure*}
\begin{center}
$\begin{array}{c}
\includegraphics[scale=0.5]{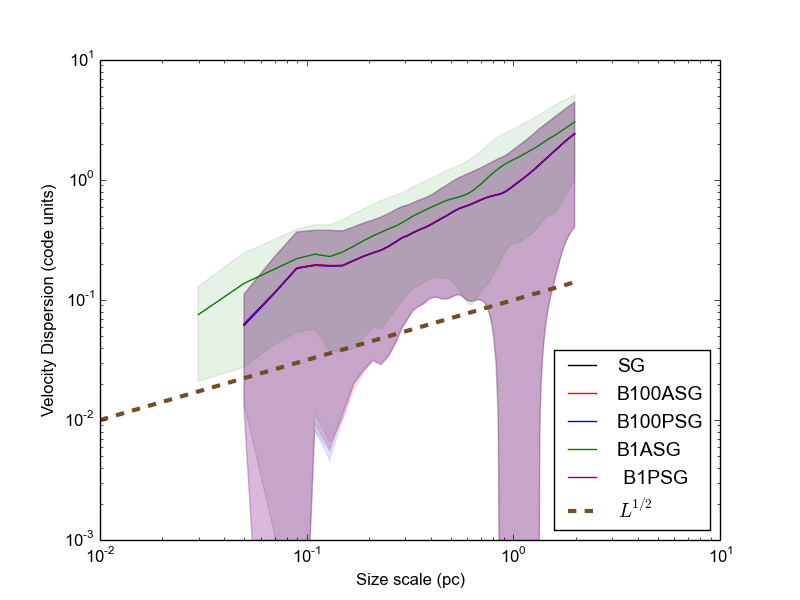} \\
\includegraphics[scale=0.5]{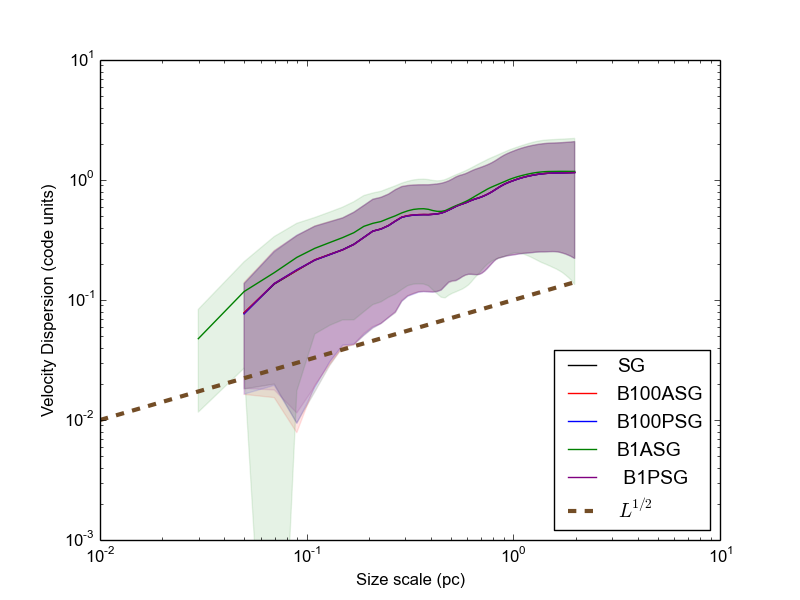} \\
\includegraphics[scale=0.5]{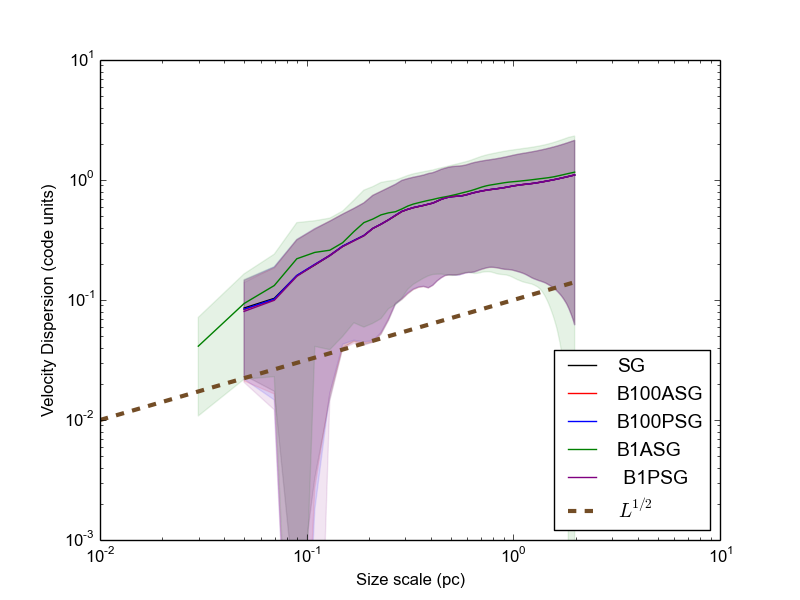} \\
\end{array}$
\end{center}
\caption{The velocity dispersion as a function of size scale for the $x$ component (top) $y$ component (middle) and $z$ component (bottom), for the ideal MHD simulations under self-gravity.  All are measured at $t=0.5$ units.}
\label{fig:SGMHD_runs}
\end{figure*}

\begin{table}
\centering
  \caption{The best-fit $\Gamma$ for the self-gravitating MHD simulations \label{tab:SGMHD_fits}}
  \begin{tabular}{c | ccc}
  \hline
  \hline
   Simulation  &  $\Gamma \,(v_x)$ & $\Gamma \,(v_y)$& $\Gamma \,(v_z)$ \\
   \hline
   B100ASG & 1.11 & 0.335 & 0.353 \\
   B100PSG  & 1.11 & 0.337 & 0.353 \\
   B1ASG & 1.06 & 0.099 & 0.301 \\
   B1PSG & 1.11 & 0.337 & 0.354 \\
 \hline
  \hline
\end{tabular}
\end{table}

\noindent Given that adding magnetic fields made little effect to the resulting $\sigma-L$ relation for non-self-gravitating gas undergoing a strong shock, we should expect a similar result for the same simulations run with self-gravity activated: i.e. that the four MHD simulations with self-gravity will resemble the control run with self-gravity (SG).  Figure \ref{fig:SGMHD_runs} shows the $\sigma-L$ relations for all five runs, and there is little difference between them.  

The slight exception is B1ASG, where we expect the $B$-field to resist compression most strongly, resulting in a slight reduction of velocity dispersion in all co-ordinates, especially in $y$ (Table \ref{tab:SGMHD_fits}).  We can see this in the middle plot of Figure \ref{fig:SGMHD_runs}, where the B1ASG curve possesses a larger scatter, and shows more velocity dispersion at $\sim 0.1-0.3$ pc scales.  Of course, this is an extreme example.  The mass-to-flux ratio of essentially all observed molecular clouds is supercritical, i.e. self-gravity will dominate over MHD effects \citep{Crutcher2012}.  Our results are in good agreement with simulations of turbulence generation through supercritical molecular cloud collisions, where clumpy structures in the collision act as efficient injectors of turbulent momentum  \citep{Wu2018}.


\subsection{Shock angle vs magnetic field angle}

\begin{figure*}
\begin{center}$
\begin{array}{cc}
\includegraphics[scale=0.4]{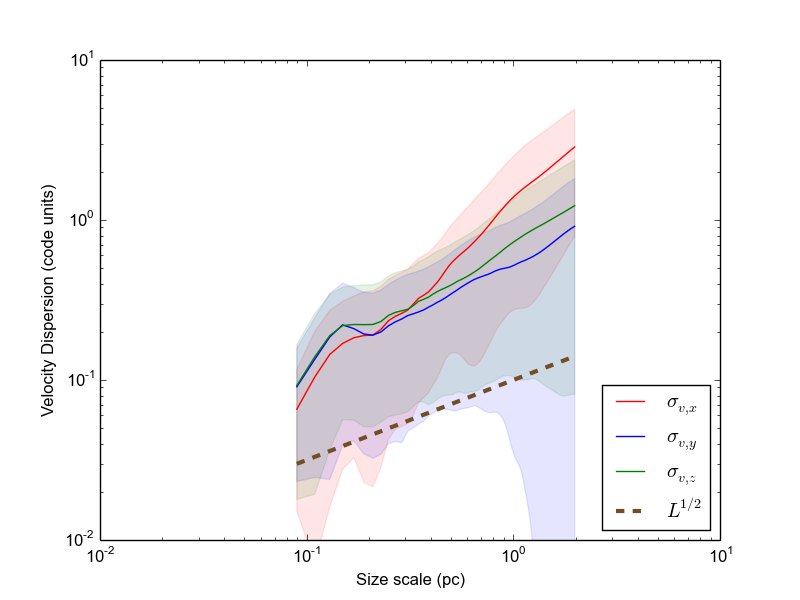} &
\includegraphics[scale=0.4]{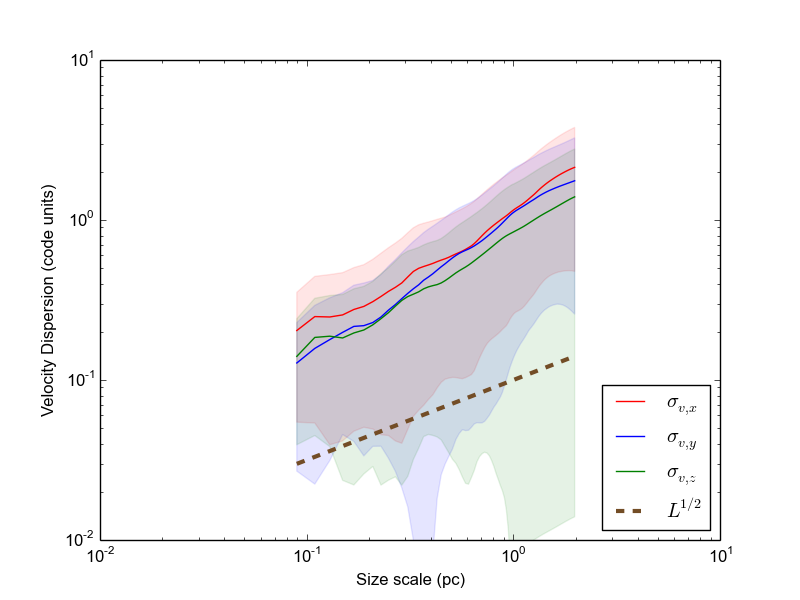} \\
\end{array}$
\end{center}
\caption{The velocity dispersion as a function of size scale for the $x,y,z$ components as a function of shock angle.  Left: MHD run with shock arriving head on, and field oriented at 45$^\circ$ (B10045); right, MHD run with field parallel to the y-axis and the shock arriving at a 45$^\circ$ angle. All are measured at $t=0.5$ units.\label{fig:MHD_angle}}
\end{figure*}

\noindent We have already seen for the pure hydrodynamic runs that modifying the shock entry angle mixes the velocity dispersion evenly between co-ordinate systems.  How does altering the direction of the field affect this mixture?

We run simulations where we keep the shock front aligned with the $y$-axis and rotate the $B$-field away by 45$^\circ$ (B100-45) and where we keep the $B$-field aligned with the $y$-axis and rotate the shock front by 45$^\circ$ (B100A-S45).  

Figure \ref{fig:MHD_angle} shows that the shock front orientation is the most important driver.  B100-45 shows similar fits to $\Gamma$ for C45, i.e. its pure hydrodynamics counterpart.  Changing the $B$-field orientation makes little appreciable effect to the fits for $\Gamma$, which are comparable to those for the B100A runs.

\section{Discussion}
\label{sec:discussion}

\noindent If clumpy spiral shocks are the driver of velocity dispersion in molecular clouds, we should expect that $\Gamma$ should vary widely over a sample of clouds, especially in spiral galaxies.  As the angle between bulk gas velocity and shock front determines $\Gamma$ along a given line of sight, we should expect that a set of molecular clouds, with a range of orientations to the spiral shock front, should produce a wide range of line of sight velocity dispersion.  This should produce significant scatter in a combined $\sigma-L$ diagram, and potentially result in no clear trend.

We note that the $\sigma-L$ trends for several nearby spiral galaxies are indeed weak - these include NGC 4526 \citep{Utomo2015} M33 \citep{Gratier2012} and M51 \citep{Colombo2014}  In the case of both M51 and NGC 4526, most clouds lie well above the $\sigma-L$ relation derived for the Milky Way \citep{Solomon1987}.  In our model, generating larger $\sigma$ across a wide range of $L$ was achieved by activating self-gravity, and was further enhanced by reducing the depth of the spiral potential (with weak magnetic fields).  

If we assume our model is correct, then we can interpret the relative differences in $\sigma-L$ relations between differing spiral galaxies as due to differences in the strength of spiral shocks, at least where they interact with molecular gas \citep[cf][]{Nguyen2018}.  More explicitly, the spiral shocks of NGC 4526 must be weaker than those of the Milky Way to be consistent with the model.  This is reasonably consistent with the fact that the molecular gas in NGC 4526 is concentrated within the inner $\sim 1$ kpc \citep{Davis2013}, where we would expect spiral shock strengths to be reduced.

The model further predicts that the relatively strong magnetic fields in the central regions of NGC 4526 will only affect the $\sigma-L$ relation if the shocks are sufficiently weak.  Otherwise, molecular cloud velocity dispersions from a sample of galaxies with varying magnetic field strength should yield no correlation between field strength and $\Gamma$.

Turbulent motions within molecular clouds should be enhanced perpendicular to the direction of recently passed shock fronts, and the level of this enhancement should depend on the time of previous passage through the shock, with those having most recently undergone passage exhibiting larger $\sigma$ at high $L$, but with an overall shallower $\Gamma$.  This is only valid for clouds that have recently passed through a shock, as magnetic braking is likely to significantly modify this, and produce quite distinct anisotropies \citep{Ossenkopf2002,Rosolowsky2003}.  Future high resolution surveys of molecular clouds in their wider environment are needed to test this prediction.

\section{Conclusions}
\label{sec:conclusions}

\noindent We have revisited the idea that passing clumpy material through spiral shocks can induce a velocity dispersion $\sigma$ vs length scale $L$ powerlaw with exponent $\Gamma \approx  0.5$.  Our work expands on previous attempts, which only considered hydrodynamic forces, and not self-gravity or magnetic fields.  The effects of these forces are tested in a suite of smoothed particle magnetohydrodynamics (SPMHD) experiments, varying the strength of the spiral shock and the strength/orientation of ambient magnetic fields.  

Every test shows an approximate powerlaw $\sigma-L$ relation with a range of values for $\Gamma=0.3-1.2$.  The key parameters that govern $\Gamma$ are the incident angle of the shock compared to the gas velocity, and the strength of self-gravity (dictated by the mass of gas entering the shock and the depth of the shock potential).  Magnetic fields show no appreciable effects, regardless of their strength and orientation relative to the shock.

We therefore conclude that velocity dispersion vs length scale powerlaws can be set up in the absence of turbulence, although obtaining exponents $\Gamma \approx 0.5$ is far from guaranteed, and depends sensitively on the geometry of the shock and the mass of gas entering it.  

If clumpy shocks are responsible for the observed $\sigma-L$ relation, this sets important constraints on the permitted shock geometry, and rough constraints on the minimum/maximum shock strengths.  If the shock is weak compared to the self-gravity of pre-shock gas, the resulting $\sigma-L$ relation will be too shallow.  Strong shocks will set up power laws that are too steep.  Magnetic fields are only effective when the shock strength is relatively weak - in this case, the velocity dispersion can be suppressed at large scales due to the addition of magnetic pressure support.

Even if clumpy shocks are not entirely responsible for the observed $\sigma-L$ relation, they remain a promising means for seeding chaotic, quasi-turbulent structures that can be later modified and enhanced by other sources of turbulence (such as instability or feedback).

\section*{Acknowledgements}

DHF and IAB gratefully acknowledge support from the ECOGAL project, grant agreement 291227, funded by the European Research Council under ERC-2011-ADG.  This  research  has  made  use  of  NASA's  Astrophysics  Data  System Bibliographic  Services.  This work relied on the compute resources of the St Andrews MHD Cluster.




\bibliographystyle{mnras} 
\bibliography{shocksinabox}


\bsp	
\label{lastpage}
\end{document}